\documentclass[aps,prd,twocolumn,superscriptaddress,nofootinbib]{revtex4-2}
\usepackage{amsmath}
\usepackage{graphicx}
\allowdisplaybreaks
\usepackage[hidelinks]{hyperref}

\begin{document}

\title{When does a sphere fall like a point particle? Quadrupole universality and Weyl-driven hexadecapole deviations in vacuum general relativity}
\author{Tiago S. Amancio}
\email{amancio@ifi.unicamp.br}
\affiliation{Instituto de F\'isica Gleb Wataghin, Universidade Estadual de Campinas, 13083-859, Campinas, S\~ao Paulo, Brazil}
\author{Ricardo A. Mosna}
\email{mosna@unicamp.br}
\affiliation{Instituto de F\'isica Gleb Wataghin, Universidade Estadual de Campinas, 13083-859, Campinas, S\~ao Paulo, Brazil}
\author{Ronaldo S. S. Vieira}
\email{ronaldo.vieira@ufabc.edu.br}
\affiliation{Centro de Ci\^encias Naturais e Humanas, Universidade Federal do ABC, 09210-580 Santo Andr\'e, SP, Brazil}

\begin{abstract}

We ask when a spinless spherical extended test body in vacuum general
relativity moves as its point-particle counterpart. In Newtonian gravity,
harmonicity of the external potential gives an all-order cancellation: in
source-free regions all spherical multipole forces beyond the monopole vanish.
Using Dixon's covariant multipole formalism, with spherical symmetry defined as
$O(3)$ invariance in the Tulczyjew-Dixon momentum rest space, we show that the
relativistic analogue holds through quadrupole order in any Ricci-flat
spacetime. At this order the torque vector vanishes, the force reduces to Ricci
contractions, and the representative worldline is geodesic; the spherical
octupole is forbidden by symmetry. This universality, however, is not an all-order
effacement principle. At hexadecapole (16-pole) order the torque vector still
vanishes, so the spinless sector remains dynamically consistent, but
curvature-squared terms generate Weyl-driven forces that can survive in
vacuum. In Schwarzschild spacetime we compute the resulting force for radial
infall and the invariant leading correction to the infall proper time. We also
show that periodic modulations of the hexadecapole moments act as an internal
drive: Melnikov's method gives transverse homoclinic splitting and local
chaotic layers near the geodesic separatrix for generic driving frequencies. The analysis is restricted to the small-body
regime of Dixon's finite multipole expansion.

\end{abstract}

\maketitle

\section{Introduction}

The motion of extended bodies in general relativity can depart from geodesic motion even in the test-body regime, due to couplings between the body's multipole moments and spacetime curvature. A systematic covariant description of these finite-size effects is provided by the Mathisson-Papapetrou-Dixon framework, in which the body's stress-energy tensor is encoded in multipole moments and the equations of motion acquire curvature-dependent force and torque terms beyond the pole-dipole approximation \cite{Mathisson1937,Papapetrou1951,Tulczyjew1959,DixonI,DixonII,DixonIII,Harte2012}. Within this perspective, it is natural to ask when a sufficiently symmetric test body moves as its point-particle counterpart, independently of its internal scalar multipole data. In relativistic language, this becomes a question of geodesic universality. The question is reminiscent of the effacement principle discussed in the post-Newtonian literature for compact bodies \cite{Damour1987,Will2014}, although the setting considered here is different: we work with extended test bodies, prescribed multipole moments, and no backreaction on the background geometry.

This broader question has a sharp benchmark in Newtonian gravity. In a source-free region the gravitational potential is harmonic, and a spherically symmetric body experiences the same net force as a point mass located at its center of mass, with vanishing torque. Consequently, in a fixed external vacuum field, Newtonian gravity realizes an all-order version of this point-particle universality: a spherical body follows the same trajectory as a point mass placed at its center of mass. The relativistic question is subtler. A Newtonian vacuum field can of course have nonzero tidal derivatives; the point is that harmonicity forces all isotropic higher-multipole contractions to collapse into Laplacians. In general relativity, Ricci-flatness ($R_{\mu\nu}=0$) likewise leaves a nonzero tidal field, encoded in the Weyl curvature, and the issue is whether spherical relativistic multipoles can couple to this trace-free curvature sector. There is no \emph{a priori} guarantee that the Newtonian shell-theorem intuition extends to relativistic finite-size effects at all multipole orders.

In recent works by the present authors, chaotic orbital motion of extended bodies was investigated in various scenarios, both in Newtonian and relativistic settings \cite{vieiraMosna2022CSF,mosnaRodriguesVieira2022PRD,rodriguesMosnaVieira2024GRG,amancioMosnaVieira2024PRD}. In the Newtonian realm, for a spinless body in the quadrupolar approximation, periodic oscillations in the body's shape were shown to trigger chaos. In the relativistic context, still at quadrupole order, spherical spinless oscillating bodies (pulsating spheres) were found to follow regular geodesic motion in Schwarzschild (and Schwarzschild-de Sitter) spacetimes; in those studies, deviations from geodesic motion for spherical bodies required backgrounds with additional structure, such as electrovacuum or ambient gravitating matter.

These examples separate three routes to finite-size chaos in fixed black-hole backgrounds. One may break spherical symmetry already at quadrupole order, allowing an anisotropic quadrupole to couple to the Weyl curvature of Schwarzschild. Alternatively, one may keep exact spherical symmetry but work in a background with a nontrivial Ricci sector, where derivatives of the Ricci curvature or of the scalar curvature allow spherical quadrupole couplings to matter or field content. The third possibility, which is the main focus here, is more intrinsically relativistic: exact spherical symmetry is retained and the background is Ricci-flat, but at sufficiently high order in the body's multipole expansion, curvature-squared couplings to the Weyl sector arise.

These results motivate the present investigation: \emph{is geodesic motion for a spherical body a generic property of vacuum spacetimes, and does it persist at arbitrary multipole order?} We address this question within Dixon's formalism for extended \emph{test} bodies \cite{DixonI,DixonII,DixonIII}, assuming the standard regime of validity of the multipole expansion: the characteristic size of the body is much smaller than the local curvature radius, and the body's backreaction on the spacetime geometry is neglected. We focus on nonrotating motion and show that this restriction is dynamically consistent for the spherical bodies considered here, because the corresponding torque vector vanishes identically in the multipole orders analyzed. To make the multipolar statement unambiguous, we formulate the calculation in the Tulczyjew-Dixon centroid convention. Unless otherwise stated, ``spherical'' therefore means $O(3)$ invariance of the multipole tensors in the instantaneous 3-space orthogonal to the linear momentum $p^\mu$, represented in an $M$-transported tetrad along the chosen worldline. This choice fixes the multipolar bookkeeping. The quadrupolar Ricci-flat cancellation itself, however, is an algebraic trace statement and not a special property of the Tulczyjew-Dixon prescription; the convention-dependence of the notion of spherical symmetry is discussed in Appendix~\ref{app:ssc_spherical}.

With these assumptions, we show that in the quadrupole approximation a spherically symmetric body is characterized by two scalar moments and has an identically vanishing torque vector; moreover, the corresponding quadrupole force reduces to Ricci-dependent terms and therefore vanishes in Ricci-flat spacetimes, implying geodesic motion for a spinless spherical body in an otherwise arbitrary vacuum spacetime (no symmetry of the background is required). In this sense, spherical quadrupole effects are Ricci-driven: the trace-free Weyl curvature drops out of the $O(3)$-invariant quadrupole coupling. We also show that, under the same symmetry requirements, the spherically symmetric octupole moment vanishes identically, so that the next nontrivial possibility occurs at the hexadecapole (16-pole) order. At this order the body is again described by two scalar moments and the torque vector still vanishes, but the multipole force may be nonzero even in Ricci-flat spacetimes; the Schwarzschild metric provides an explicit example. The transition from quadrupole to hexadecapole order is therefore qualitative rather than merely quantitative: curvature-squared terms allow Weyl-driven finite-size effects in vacuum. We then present two applications in Schwarzschild spacetime: (i) we compute the leading hexadecapolar correction to the proper time of radial infall from a finite radius; (ii) allowing periodic modulations of the hexadecapole moments (pulsations), the equatorial dynamics becomes a time-periodically forced Hamiltonian system, and Melnikov's method is used to identify chaotic layers near the separatrix associated with the unstable circular orbit.

This paper is organized as follows. We begin, in Sec.~\ref{secnewton}, with a summary of the Newtonian results regarding the orbit of a spherical test body in vacuum, in order to provide a comparison with the relativistic case. Section~\ref{dixon} reviews the general aspects of Dixon's formalism. We present the results concerning spherical test bodies in the quadrupole and hexadecapole approximations in Secs.~\ref{sec4pole} and~\ref{sec16pole}, respectively. Sec.~\ref{secfall} applies the formalism to calculate the hexadecapole corrections for the proper time of radial infall of a spinless sphere in Schwarzschild spacetime. In Sec.~\ref{secchaos}, we apply Melnikov's method to identify chaotic dynamics in the motion of pulsating spheres in the equatorial plane of Schwarzschild spacetime at hexadecapole order. The final conclusions are summarized in Sec.~\ref{conclusion}. We adopt the metric signature ``$-+++$'' and the definitions of the Riemann tensor as $\nabla_\mu\nabla_\nu\omega_\lambda-\nabla_\nu\nabla_\mu\omega_\lambda={R_{\mu\nu\lambda}}^\rho\omega_\rho$, and of the Ricci tensor as $R_{\mu\nu}={R_{\mu\lambda\nu}}^\lambda$.

\section{Spherical test bodies in Newtonian gravity}
\label{secnewton}

Consider a localized but otherwise arbitrary mass distribution generating a Newtonian gravitational potential $\Phi$. In the source-free region exterior to this mass distribution, the potential is harmonic, i.e.\ it satisfies Laplace's equation,
\begin{equation}
\partial^a \partial_a \Phi = 0.
\end{equation}
We now place, in this external field, a sufficiently small extended body with mass density $\rho(\mathbf{x},t)$. Its total mass and center-of-mass position are defined as
\begin{align}
  m &= \int_{\mathcal{R}} \rho(\mathbf{x},t)\,d^3\mathbf{x}, \\
  \mathbf{z}(t) &= \frac{1}{m}\int_{\mathcal{R}} \mathbf{x}\,\rho(\mathbf{x},t)\,d^3\mathbf{x},
  \label{ncom}
\end{align}
where the (possibly time-dependent) region of integration $\mathcal{R}$ is the topologically compact support of the mass density $\rho$.

The total force and the torque (relative to the center of mass) experienced by this body are
\begin{align}
  \mathbf{F}(t) &= -\int_{\mathcal{R}} \rho(\mathbf{x},t)\,\nabla\Phi\,d^3\mathbf{x},
  \label{nforce}\\
  \mathbf{N}(t) &= -\int_{\mathcal{R}} \rho(\mathbf{x},t)\,(\mathbf{x}-\mathbf{z}(t))\times\nabla\Phi\,d^3\mathbf{x}.
  \label{ntorque}
\end{align}
By commutativity of partial derivatives, if $\Phi$ is harmonic then each component of its gradient is also harmonic:
\begin{equation}
\partial^a\partial_a(\partial_i\Phi)=\partial_i\partial^a\partial_a\Phi=0.
\end{equation}
Thus, to compute the force $\mathbf{F}$ we may use the mean-value theorem for harmonic functions, which states that the mean value of a harmonic function $f$ over a sphere is equal to its value at the center:
\begin{equation}
  f(\mathbf{x}_0)=\frac{1}{4\pi R^2}\int_{S_R(\mathbf{x}_0)} f(\mathbf{x})\,dS,
  \label{meanvalue}
\end{equation}
where $S_R(\mathbf{x}_0)$ denotes the sphere of radius $R$ centered at $\mathbf{x}_0$.

If the extended body is spherically symmetric, its mass density $\rho$ depends only on the norm of the separation vector from the center of mass, $\mathbf{r}=\mathbf{x}-\mathbf{z}$. Evaluating the average of each component of $\nabla\Phi$ over spheres centered at $\mathbf{z}$ then gives the well-known result
\begin{equation}
  F_i = -m\,\partial_i\Phi(\mathbf{z}).
\end{equation}
In other words, the total force on the body is the same as that on a point particle of mass $m$ located at its center of mass. Since in Newtonian gravity the orbital motion and the body's spin do not couple, this already implies that a spherical body in vacuum follows the same orbit as a point particle with the same initial center-of-mass position and velocity.

The torque can also be simplified. A vector identity allows us to rewrite Eq.~\eqref{ntorque} as
\begin{equation}
  \mathbf{N} = \int_{\mathcal{R}}\left\{\nabla\times\left[\rho\,\Phi\,\mathbf{r}\right]
  - \Phi\,\nabla\times\left[\rho\mathbf{r}\right]\right\}\,d^3\mathbf{x}.
\end{equation}
The first term can be converted into the surface integral
\begin{equation}
  -\oint_{\partial\mathcal{R}} \rho\,\Phi\,\mathbf{r}\times d\mathbf{S},
\end{equation}
which vanishes when evaluated on a spherical boundary $\partial\mathcal{R}$, since $\mathbf{r}$ and $d\mathbf{S}$ are parallel. The second term simplifies to
\begin{equation}
  \mathbf{N} = -\int_{\mathcal{R}} \Phi\,\nabla\rho\times\mathbf{r}\,d^3\mathbf{x}.
  \label{ntorquesimp}
\end{equation}
Imposing again spherical symmetry, $\rho=\rho(r)$, where $r=|\mathbf{r}|$, so that $\nabla\rho=\rho'(r)\hat{\mathbf{r}}$ is radial, the integrand in Eq.~\eqref{ntorquesimp} then vanishes pointwise, and we conclude that the torque on a spherical body is zero. Notice that this conclusion does \emph{not} rely on the harmonicity of $\Phi$: within the scalar-potential force law assumed above, the vanishing of $\mathbf{N}$ is purely a consequence of spherical symmetry and does not depend on whether the background is vacuum or on any symmetry of the external field.

For later comparison with the relativistic analysis, it is useful to rephrase these results in multipolar language. Expanding the potential around the body's center of mass and introducing the raw multipole moments of the mass distribution, one finds that the total force and torque can be written as a monopole term plus a series of higher multipole contributions. For a spherically symmetric body in a harmonic potential, isotropy implies that each even-order moment is built from products of Kronecker deltas, and every contraction with derivatives of $\Phi$ produces Laplacians, which vanish by $\partial^a\partial_a\Phi=0$. As shown in Appendix~\ref{ap:newtexp}, this leads to a complete ``effacement'' of all multipole orders beyond the monopole: each contribution to the force and torque above monopole vanishes separately, and the motion of a spherical body in Newtonian vacuum is exactly that of a point particle. In the next section we investigate to what extent this Newtonian universality persists for small spherical test bodies in general relativity.

\section{Dixon's formalism for spherically symmetric bodies}
\label{dixon}

Having reviewed the Newtonian universality of free fall for spherical bodies in vacuum, we now address the corresponding question in general relativity for \emph{small extended test bodies}. Our analysis is based on Dixon's covariant multipole formalism \cite{DixonI,DixonII,DixonIII}, which provides equations of motion for the body's linear and angular momenta in terms of curvature-dependent force and torque contributions.

\subsection*{Assumptions and conventions}

We consider an extended body whose backreaction on the background spacetime $(\mathcal{M},g)$ is negligible (test-body approximation). Moreover, we assume that the characteristic linear size of the body is much smaller than the local curvature radius along its world tube, so that a multipole expansion is meaningful. The body is described by a stress-energy tensor $T^{\mu\nu}$ satisfying
\begin{align}
  \nabla_\mu T^{\mu\nu}=0.
\end{align}

Dixon's construction associates to the body a reference worldline $z^\mu(s)$ (with arbitrary parameter $s$), a linear momentum $p^\mu(s)$, and a spin tensor $S^{\mu\nu}(s)$ whose evolution is governed by
\begin{align}
  \frac{Dp^\mu}{ds}&=-\frac{1}{2}{R^\mu}_{\nu\alpha\beta}\,v^\nu S^{\alpha\beta}+F^\mu,
  \label{dpdixon} \\
  \frac{DS^{\mu\nu}}{ds}&=2p^{[\mu}v^{\nu]}+N^{\mu\nu},
  \label{dsdixon}
\end{align}
where $v^\mu\equiv dz^\mu/ds$, ${R^\mu}_{\nu\alpha\beta}$ is the Riemann tensor, and $F^\mu$ and $N^{\mu\nu}$ are the multipole force and torque relative to the chosen worldline. We adopt the Tulczyjew-Dixon spin supplementary condition
\begin{equation}
  S^{\mu\nu}(s)\,p_{\nu}(s) = 0,
  \label{centerofmass}
\end{equation}
which selects $z^\mu(s)$ as a covariant center-of-mass worldline.

The mass $m$ associated with the body is defined by
\begin{align}
  p_\mu p^\mu=-m^2,
  \label{defmass}
\end{align}
and is not constant in general. Writing $p^\mu = m u^\mu$ defines a unit timelike vector $u^\mu$ parallel to $p^\mu$. In general $u^\mu$ and $v^\mu$ need not be parallel.

\subsection*{Multipole force and torque; tensor extensions}

The explicit forms of $F^\mu$ and $N^{\mu\nu}$ in Eqs.~\eqref{dpdixon}--\eqref{dsdixon} are given in terms of multipole moments $I^{\sigma_1 \dots\sigma_n\alpha\beta}$ and tensor extensions of the metric \cite{DixonIII}:
\begin{align}
  F^{\mu}&=\frac{1}{2}\sum_{n=2}^{\infty}\frac{1}{n!}I^{\sigma_1 \dots\sigma_n\alpha\beta}\,
  \nabla^{\mu}g_{\alpha\beta,\sigma_1 \dots\sigma_n}\big(z\big),
  \label{generalforce}\\
  N^{\mu\nu}&=\sum_{n=1}^{\infty}\frac{1}{n!}\,g^{\lambda[\mu}I^{\nu]\sigma_1 \dots\sigma_n\alpha\beta}\,
  g_{\{\lambda\alpha,\beta\}\sigma_1 \dots\sigma_n}\big(z\big),
  \label{generaltorque}
\end{align}
where $A_{\{\alpha\beta\gamma\}}=A_{\alpha\beta\gamma}-A_{\beta\gamma\alpha}+A_{\gamma\alpha\beta}$.

The quantity $g_{\alpha\beta,\sigma_1\dots\sigma_n}$ denotes the $n$th tensor extension of the metric. In a Riemann normal coordinate system $\{y^a\}$ centered at the point $z$ (the pole), tensor extensions are defined by
\begin{align}
  g_{ab,c_1\dots c_n}\big(z\big)=
  \left.\frac{\partial}{\partial y^{c_1}}\dots\frac{\partial}{\partial y^{c_n}}g_{ab}\right|_{z}.
\end{align}
Latin indices refer to components in such normal-coordinate frames. Tensor extensions provide a covariant Taylor expansion of tensor fields in a normal neighborhood; see \cite{DixonIII,harte} for details. We will need the explicit identities \cite{harte}
\begin{align}
  g_{\alpha\beta,\sigma_1 \sigma_2}=&\frac{2}{3}R_{\alpha(\sigma_1 \sigma_2)\beta},
  \label{g4def}\\
  g_{\alpha\beta,\sigma_1 \sigma_2 \sigma_3 \sigma_4}=&\frac{6}{5}\nabla_{(\sigma_1 \sigma_2}R_{|\alpha|\sigma_3 \sigma_4)\beta}
  +\frac{16}{15}{R_{\alpha(\sigma_1 \sigma_2}}^{\lambda}R_{|\beta|\sigma_3 \sigma_4)\lambda},
  \label{g6def}
\end{align}
as well as
\begin{align}
  g_{(\alpha\beta),\sigma_1\dots\sigma_n}=g_{\alpha\beta,\sigma_1\dots\sigma_n}=g_{\alpha\beta,(\sigma_1\dots\sigma_n)},\\
  g_{\alpha(\beta,\sigma_1\dots\sigma_n)}=g_{(\alpha\beta,\sigma_1\dots\sigma_{n-1})\sigma_n}=0.
\end{align}

\subsection*{Riemann-type multipoles, spin/torque vectors, and conserved quantities}

The tensors $I^{\sigma_1\dots\sigma_n\alpha\beta}$ are the multipole moments of $T^{\alpha\beta}$: the quadrupole is $I^{\sigma_1\sigma_2\alpha\beta}$, the octupole is $I^{\sigma_1\sigma_2\sigma_3\alpha\beta}$, etc. The monopole and dipole information is encoded in $p^\mu$ and $S^{\mu\nu}$. Importantly, once a hierarchy of multipole moments is specified, Eqs.~\eqref{dpdixon}--\eqref{dsdixon} provide evolution equations for $p^\mu$ and $S^{\mu\nu}$; the higher multipoles are not fixed by the equations of motion alone and must be supplied by an internal model (or treated as prescribed data) within the regime of validity of the expansion.

It is convenient to introduce moments $J^{\sigma_1 \dots\sigma_n\alpha\beta\mu\nu}$ with Riemann-type index symmetries, defined by
\begin{align}
  J^{\sigma_1\dots\sigma_n\alpha\beta\mu\nu}
  =I^{\sigma_1\dots\sigma_n[\alpha[\mu\beta]\nu]}, \quad n \geq 0,
\end{align}
where
$A^{[\mu[\nu\alpha]\beta]}=\frac{1}{4}\left(A^{\mu\nu\alpha\beta}-A^{\alpha\nu\mu\beta}-A^{\mu\beta\alpha\nu}+A^{\alpha\beta\mu\nu}\right)$.
These moments satisfy
\begin{align}
  &J^{(\sigma_1\dots\sigma_n)[\alpha\beta][\mu\nu]}=J^{\sigma_1\dots\sigma_n\alpha\beta\mu\nu}, \quad n\geq 0,
  \label{jprop1}\\
  &J^{\sigma_1\dots\sigma_n\alpha[\beta\mu\nu]}=0, \quad n\geq 0,
  \label{jprop2}\\
	& J^{\sigma_1\dots\sigma_{n-1}[\sigma_n \alpha \beta] \mu \nu}=0, \quad n\geq 1,
	\label{jprop3}\\
  &u_{\sigma_1} J^{\sigma_1\dots\sigma_n\alpha\beta\mu\nu}=0, \quad n\geq 1.
  \label{jprop4}
\end{align}
The last four indices have the algebraic symmetries of the Riemann tensor.

Given the spin supplementary condition \eqref{centerofmass}, it is often useful to work with the spin vector
\begin{align}
  S_\mu=-\frac{1}{2}\epsilon_{\mu\nu\alpha\beta}u^\nu S^{\alpha\beta},
  \label{spinvector}
\end{align}
where $\epsilon_{\mu\nu\alpha\beta}$ is the Levi-Civita tensor (we take $\epsilon_{0123}=+\sqrt{-g}$, equivalently $\epsilon_{0123}=+1$ in a local orthonormal frame). This can be inverted as
\begin{align}
  S^{\mu\nu}=g^{\mu\kappa}g^{\nu\lambda}\epsilon_{\kappa\lambda\alpha\beta}u^\alpha S^\beta.
  \label{spinvectorinv}
\end{align}
Similarly, we define the torque vector
\begin{align}
  N_\mu=-\frac{1}{2}\epsilon_{\mu\nu\alpha\beta}u^\nu N^{\alpha\beta},
  \label{torquevector}
\end{align}
This definition is only a projection of $N^{\mu\nu}$: unlike $S^{\mu\nu}$, the antisymmetric tensor $N^{\mu\nu}$ cannot be reconstructed from $N^\mu$ alone. Combining these definitions with Eq.~\eqref{dsdixon} yields the useful relation
\begin{align}
  N^\mu=\frac{DS^\mu}{ds}-u^\mu\frac{Du_\nu}{ds}S^\nu.
  \label{spinvectorevo}
\end{align}

Finally, for the exact Dixon equations, each Killing vector field $\xi^\mu$ of $(\mathcal{M},g)$ defines the conserved generalized momentum \cite{DixonIII,ehlersRudolph1977GRG}
\begin{align}
  \mathcal{P}_{\xi}=p_\mu\xi^\mu+\frac{1}{2}S^{\mu\nu}\nabla_\mu\xi_\nu.
  \label{conserved}
\end{align}
The same expression is preserved in a finite truncation provided the force and torque are truncated consistently. Higher multipoles do not appear explicitly in $\mathcal{P}_{\xi}$; rather, their force and torque contributions satisfy the Killing-vector identity required for $d\mathcal{P}_{\xi}/ds=0$ through the order retained.

\subsection{Spherically symmetric bodies in the quadrupole approximation}
\label{sec4pole}

We now specialize to \emph{spherically symmetric} bodies. In the Tulczyjew-Dixon construction adopted here, this means $O(3)$ invariance in the instantaneous 3-space orthogonal to the momentum $p^\mu$. Equivalently, using $p^\mu=m u^\mu$ with $u_\mu u^\mu=-1$, we impose isotropy in $u^\perp=p^\perp$. Operationally, we represent the multipoles in a nonrotating, momentum-adapted frame furnished by Dixon's $M$-transport \cite{DixonIII}. A vector field $A^\kappa$ along the worldline is $M$-transported if
\begin{align}
  \frac{DA^\kappa}{ds}-\left(u^\kappa\frac{Du_\lambda}{ds}-\frac{Du^\kappa}{ds}u_\lambda\right)A^\lambda=0.
\end{align}
$M$-transport furnishes a notion of ``nonrotation'' along the worldline. In particular, one may choose an $M$-transported orthonormal tetrad $\{e_a\}$ with $e_0=u$ and $\{e_i\}_{i=1,2,3}$ spanning the 3-space orthogonal to $u^\mu$ (equivalently to $p^\mu$). This tetrad provides the nonrotating, momentum-adapted frame used below. The dependence of this notion of spherical symmetry on the centroid/rest-space prescription is discussed in Appendix~\ref{app:ssc_spherical}.

In terms of $J^{\alpha\beta\gamma\delta}$, the quadrupole force and torque can be written as
\begin{align}
  F^\mu=&-\frac{1}{6}J^{\alpha\beta\gamma\delta}\nabla^\mu R_{\alpha\beta\gamma\delta},
  \label{forcequad}\\
  N^{\mu\nu}=&\frac{4}{3}J^{\alpha\beta\gamma[\mu}{R^{\nu]}}_{\gamma\alpha\beta}.
  \label{torquequad}
\end{align}
In the $M$-transported frame $\{e_a\}$, spherical symmetry implies that the components $J^{abcd}$ are invariant under transformations acting on the spatial indices $i=1,2,3$. A standard representation theorem for isotropic tensors states that $SO(3)$-invariant tensors are generated by products of $\delta_{ij}$ and, possibly, one factor of $\epsilon_{ijk}$. Requiring invariance under the full group $O(3)$ excludes the parity-odd $\epsilon_{ijk}$ sector, leaving products of Kronecker deltas \cite{Jeffreys1973}. Together with the Riemann-type symmetries in Eqs.~\eqref{jprop1} and \eqref{jprop2}, this leaves only two independent nonzero components of $J^{abcd}$,
\begin{align}
  j_{m}:=J^{0101}=J^{0202}=J^{0303},\\
  j_{s}:=J^{2323}=J^{1313}=J^{1212},
\end{align}
so that the quadrupole tensor is fully specified by two scalars $j_m$ and $j_s$, interpreted as mass and stress quadrupole moments \cite{ehlersRudolph1977GRG}. Reconstructing $J^{\alpha\beta\gamma\delta}$ from these scalars yields a particularly transparent form if one introduces the spatial projector onto the momentum rest space,
\begin{align}
  h^{\mu\nu}=g^{\mu\nu}+u^\mu u^\nu.
  \label{hprojector}
\end{align}
In terms of $h^{\mu\nu}$, the spherical quadrupole is
\begin{align}
  J^{\alpha\beta\gamma\delta}
  =&2j_s h^{\alpha[\gamma}h^{\delta]\beta}
  -4j_m u^{[\alpha}h^{\beta][\gamma}u^{\delta]}.
  \label{jquad_h}
\end{align}
This form makes explicit that the two independent pieces are built only from $u^\mu$ and the isotropic spatial metric $h^{\mu\nu}$ on the momentum 3-space. Expanding $h^{\mu\nu}=g^{\mu\nu}+u^\mu u^\nu$ gives the equivalent compact expression
\begin{align}
  J^{\alpha\beta\gamma\delta}
  =&2j_s g^{\alpha[\gamma}g^{\delta]\beta}
  -4(j_m+j_s)u^{[\alpha}g^{\beta][\gamma}u^{\delta]}.
  \label{jquad}
\end{align}

From Eq.~\eqref{torquequad} one obtains
\begin{align}
  N^{\mu\nu}=\frac{8}{3}(j_m+j_s){R^{[\mu}}_{\lambda}\, u^{\nu]}u^\lambda,
  \label{sphtorque}
\end{align}
so that $N^{\mu\nu}=0$ in vacuum ($R_{\mu\nu}=0$). Moreover, substituting \eqref{sphtorque} into \eqref{torquevector} shows that the torque vector $N_\mu$ vanishes \emph{identically} for any spherically symmetric body in the quadrupole approximation, irrespective of whether the background is vacuum or not. Therefore, if $S^\mu(s_0)=0$ at some instant, Eq.~\eqref{spinvectorevo} implies that $S^\mu$ remains zero, and hence $S^{\mu\nu}=0$ is a consistent solution of Eq.~\eqref{dsdixon} in the quadrupole approximation.

Substituting Eq.~\eqref{jquad} into Eq.~\eqref{forcequad} yields
\begin{align}
  F_\mu=-\frac{1}{3}j_s\nabla_\mu R-\frac{2}{3}(j_m+j_s)u^\alpha u^\beta \nabla_\mu R_{\alpha\beta},
  \label{quadforcesph}
\end{align}
where $R$ is the Ricci scalar. Equations~\eqref{sphtorque} and \eqref{quadforcesph} fit naturally into the general vacuum reduction of Dixon's quadrupole: the quadrupolar force and torque depend only on its four-dimensional trace-free part \cite{Harte2020,HarteDwyer2023}. For the spherical tensor \eqref{jquad_h}, exact $O(3)$ invariance sets that part to zero, leaving only the Ricci-sector terms displayed above. Thus the Weyl curvature cannot drive a spherical quadrupole, and in Ricci-flat spacetimes
\[
  F_\mu=0, \qquad N^{\mu\nu}=0.
\]
If in addition $S^{\mu\nu}=0$, Eq.~\eqref{dpdixon} reduces to
\begin{align}
  \frac{Dp^\mu}{ds}=0.
  \label{dpszero}
\end{align}
Geodesic motion follows in the standard way. Choosing $s=\tau$ as the proper time along the center-of-mass worldline, $v_\mu v^\mu=-1$. With $S^{\mu\nu}=0$ and $N^{\mu\nu}=0$, Eq.~\eqref{dsdixon} implies $p^\mu v^\nu=p^\nu v^\mu$, hence $p^\mu\propto v^\mu$, and therefore $v^\mu=u^\mu$. Since $p^\mu=m u^\mu$, Eq.~\eqref{dpszero} further implies $Dm/d\tau=0$, and the equation of motion reduces to
\begin{align}
  \frac{Dv^\mu}{d\tau}=0,
  \label{eq:DvmuDs0}
\end{align}
i.e.\ the center-of-mass worldline is geodesic.

\subsection{Spherically symmetric body in the hexadecapole approximation}
\label{sec16pole}

We now proceed to the next nontrivial multipole order compatible with spherical symmetry. Under the algebraic symmetries \eqref{jprop1}--\eqref{jprop4} together with $O(3)$ invariance in the $p^\mu$-orthogonal 3-space, the octupole moment of a spherical body vanishes identically. Consequently, the next nonzero moment is generically the hexadecapole tensor $J^{\alpha\beta\gamma\delta\mu\nu}$.

Using the symmetry properties of $J^{\alpha\beta\gamma\delta\mu\nu}$ and the tensor-extension identity \eqref{g6def}, the hexadecapole contributions to the force $F_\mu$ and torque $N^{\mu\nu}$ in Eqs.~\eqref{generalforce} and \eqref{generaltorque} can be rewritten, for an arbitrary body, as
\begin{align}
  F_\mu=&\,J^{\alpha\beta\gamma\delta\rho\sigma}\nabla_\mu \left(
  \frac{1}{20} \nabla_{\alpha}\nabla_{\beta} R_{\rho\gamma\delta\sigma}
  -\frac{4}{45} {R_{\rho\alpha\beta}}^\lambda R_{\sigma\gamma\delta\lambda}\right),
  \label{16force}\\
  N^{\mu\nu} = &\,\frac{2}{5} J^{\alpha\beta\gamma\rho\sigma[\mu} g^{\nu]\delta}\Big(
  \nabla_{\gamma}\nabla_\beta R_{\alpha\rho\delta\sigma}
  + \nabla_{(\gamma}\nabla_{\delta)} R_{\alpha\rho\beta\sigma}
  \nonumber\\
  &\hspace{6.0em}
  -\frac{16}{9} {R_{\alpha\rho\beta}}^\lambda R_{(\gamma|\sigma|\delta)\lambda}\Big).
  \label{16torque}
\end{align}
The derivation of these expressions is algebraically involved and was checked with the aid of symbolic computation using Mathematica~\cite{mathematica}.

Imposing spherical symmetry in the $M$-transported frame $\{e_a\}$, one finds that the only independent nonzero components of $J^{abcdef}$ are
\begin{align}
  J_m:=J^{110101},\\
  J_s:=J^{111313}.
\end{align}
These scalars completely determine the hexadecapole tensor and can be interpreted as mass and stress hexadecapole moments. For example, in Minkowski spacetime, for a spherical perfect fluid characterized by mass density $\rho$ and pressure $P$, one has
\begin{align}
  J_m = \int 3\rho\, x^4\,dV,
  \qquad
  J_s = \int P\, x^2\left(x^2+z^2\right)\,dV,
  \label{integralJ}
\end{align}
in suitable Cartesian coordinates, where $dV$ is the Euclidean volume element. We note that, by isotropy, any component choice is equivalent up to numerical factors; the expressions above illustrate positivity/scaling.

Reconstructing the covariant tensor $J^{\alpha \beta \gamma \delta \mu \nu }$ from these components yields
\begin{align}
  J^{\alpha \beta \gamma \delta \mu \nu }=&
  J_{(A)}^{(\alpha \beta) [\gamma \delta] [\mu \nu] }
  +J_{(B)}^{(\alpha \beta) [\gamma \delta] [\mu \nu] }
  \nonumber\\
  &+J_{(C)}^{(\alpha \beta) [\gamma \delta] [\mu \nu] }
  +J_{(D)}^{(\alpha \beta) [\gamma \delta] [\mu \nu] },
  \label{J16}
\end{align}
with
\begin{align}
  J_{(A)}^{\alpha \beta \gamma \delta \mu \nu }=&\,
  J_s \left(2 g^{\alpha\mu}g^{\beta\gamma}g^{\delta\nu}+ g^{\alpha\beta}g^{\gamma\mu}g^{\delta\nu}\right),\\
  J_{(B)}^{\alpha \beta \gamma \delta \mu \nu }=&\,
  4 J_m g^{\alpha(\beta}g^{\gamma\mu)}u^\delta u^\nu,\\
  J_{(C)}^{\alpha \beta \gamma \delta \mu \nu }=&\,
  J_s \left(u^\alpha u^\beta g^{\gamma\mu} g^{\delta\nu}
  + 4 g^{\alpha(\beta}g^{\gamma)\mu}u^\delta u^\nu \right. \nonumber\\ & \left.+  4 g^{\alpha(\delta}u^{\nu)}u^\beta g^{\gamma\mu}\right),\\
  J_{(D)}^{\alpha \beta \gamma \delta \mu \nu }=&\,
  \frac{4}{3} \left(J_m +3 J_s\right) u^\alpha u^\beta g^{\gamma\mu} u^\delta u^\nu.
\end{align}

Substituting Eq.~\eqref{J16} into Eq.~\eqref{16torque} and then using Eq.~\eqref{torquevector} shows that the torque vector $N^\mu$ vanishes identically for a spherical body (although $N^{\mu\nu}$ need not vanish). Therefore, as in the quadrupole case, a configuration with $S^{\mu\nu}=0$ at some instant remains spinless in the hexadecapole approximation.

The force behaves differently. In contrast with Eq.~\eqref{quadforcesph}, Eq.~\eqref{16force} contains not only derivatives of curvature but also terms quadratic in the Riemann tensor. In Ricci-flat spacetimes the Riemann tensor is the Weyl tensor, so these curvature-squared terms provide a genuinely Weyl-driven channel for finite-size effects. As a consequence, the hexadecapole force \emph{may be nonzero even in Ricci-flat spacetimes}. Hence, already at this next nontrivial multipole order, spherical symmetry is not sufficient to guarantee geodesic motion in vacuum; cancellations can still occur for special choices of the material parameters $J_m$ and $J_s$, but they are not enforced by spherical symmetry alone. In the next sections we exhibit explicit nonzero hexadecapole forces in Schwarzschild spacetime, considering radial infall and time-dependent (pulsating) spherical bodies.

\section{Application: Radial infall in (vacuum) Schwarzschild spacetime}
\label{secfall}

The results of Sec.~\ref{sec16pole} show that, at hexadecapole order, a spinless spherically symmetric test body may experience a nonvanishing multipole force even in Ricci-flat spacetimes. In this first application we exhibit this effect explicitly in Schwarzschild spacetime, the canonical Ricci-flat black-hole geometry, and quantify it through a clean invariant measure: the correction to the proper time of radial infall.

The Schwarzschild line element is
\begin{align}
  ds^2=-f(r)\,dt^2+f(r)^{-1}dr^2+r^2\left(d\theta^2+\sin^2\theta\,d\phi^2\right),
\end{align}
where $f(r)=1-2M/r$ and $M$ is the black-hole mass.

As discussed in Sec.~\ref{sec16pole}, a spinless configuration satisfies Eq.~\eqref{dsdixon}. Choosing the evolution parameter $s$ to be the proper time $\tau$ along the center-of-mass worldline, Eq.~\eqref{dpdixon} reduces to
\begin{align}
  \frac{Dp_\mu}{d\tau}=F_\mu,
  \label{eqmotionsimp}
\end{align}
where $F_\mu$ is the hexadecapole force in Eq.~\eqref{16force} specialized to a spherically symmetric body via Eq.~\eqref{J16}.

Schwarzschild spacetime admits the Killing vectors $\partial_t$ and $\partial_\phi$. For the consistently truncated spinless hexadecapole system used here, Eq.~\eqref{conserved} reduces, through the order retained, to the two conserved quantities
\begin{align}
  E \equiv -p_t, \qquad L \equiv p_\phi,
\end{align}
which we interpret as energy and angular momentum. For radial infall we impose $\theta=\pi/2$, $u^\theta=0$, and $L=0$. Since $p^\mu=m u^\mu$, the condition $L=0$ implies $u^\phi=0$.

With these restrictions, the hexadecapole torque tensor computed from Eqs.~\eqref{16torque} and~\eqref{J16} vanishes, and therefore $p^\mu \parallel v^\mu$, as before; hence $v^\mu=u^\mu$. Moreover, the hexadecapole force has a single nonzero component,
\begin{align}
  F_r = \frac{h M^2}{r^7},
	\qquad
  h \equiv \frac{4}{15}\left(8 J_m - 21 J_s\right).
\end{align}
Equation~\eqref{eqmotionsimp} then reduces to two independent equations:
\begin{align}
  \dot{t} = \frac{E r^2}{(r-2M)^2 p_r}\,\dot{r},
  \label{dott}\\
  \dot{p}_r + \frac{M p_r}{r(r-2M)}\dot{r} + \frac{M E}{r(r-2M)} \dot{t} = F_r,
\end{align}
where a dot denotes differentiation with respect to $\tau$. (The remaining components confirm \emph{a posteriori} that $\theta=\pi/2$ and $u^\theta=0$ are consistent.)

Using $p_r=m u_r=m v_r$, one eliminates $p_r$ from Eq.~\eqref{dott} and substitutes both $p_r$ and $\dot{t}$ into the normalization $v_\mu v^\mu=-1$, obtaining
\begin{align}
  \dot{r}^2 = \frac{E^2}{m^2}-\left(1-\frac{2M}{r}\right).
  \label{eqradfall}
\end{align}
To determine the variation of $m$, we write $p_\mu=m v_\mu$ in Eq.~\eqref{eqmotionsimp} and contract with $v^\mu$, which gives
\begin{align}
  \dot{m} = - \frac{h M^2}{r^7}\,\dot{r}.
\end{align}
Integrating,
\begin{align}
  m(r) = E + \frac{h M^2}{6 r^6},
\end{align}
where the integration constant was chosen equal to $E$, so that the motion is compatible with infall from rest at infinity and $E=m_0$ in the asymptotically flat region. Substituting $m(r)$ back into Eq.~\eqref{eqradfall} yields
\begin{align}
  \dot{r}^2 =
  \left(1+\frac{h M^2}{6 E r^6}\right)^{-2}-\left(1-\frac{2M}{r}\right).
  \label{eqradfall2}
\end{align}
Equation~\eqref{eqradfall2} can be integrated to obtain the proper time of infall. Although the trajectory can be thought of as originating from rest at infinity, the total proper time from infinity diverges. We therefore compute the proper time accumulated between a finite radius $R$ and the horizon $r=2M$. Within the regime of validity of the multipole expansion, the hexadecapole parameters are necessarily small, so we expand Eq.~\eqref{eqradfall2} to first order in $h$. The resulting proper time is
\begin{align}
  \Delta \tau =\,&\frac{4M}{3}\left[\left(\frac{R}{2M}\right)^{3/2}-1\right]
  \nonumber\\
  &+ \frac{h}{672 E M^3}\left[1-\left(\frac{R}{2M}\right)^{-7/2}\right].
\end{align}
The first term is the familiar point-particle result (see, e.g., \cite{schutz}). The bracket multiplying $h$ in the correction term is positive for any $R>2M$, so the sign of the hexadecapole correction is controlled entirely by $h$.

This provides a simple physical reading of the effect. The hexadecapole force scales as $F_r\sim r^{-7}$ and is therefore strongly suppressed at large radii; nevertheless it produces a cumulative correction to the infall proper time. This correction is proportional to $h$ and its magnitude is bounded by $|h|/(672 E M^3)$. Comparing with a point particle released from rest at infinity and measuring the proper time over the interval from $R$ to $2M$, the spherical body takes a longer proper time to reach the horizon if $h>0$, whereas $h<0$ would imply a shorter infall time. For simple matter models (e.g.\ perfect fluids with $P\ll \rho$), Eq.~\eqref{integralJ} suggests that $J_s$ is typically suppressed relative to $J_m$, so that $h$ is expected to be positive, and the extended spherical body falls slightly ``slower'' than a point particle even in vacuum Schwarzschild spacetime. This explicitly illustrates the breakdown of point-particle universality at the first nontrivial spherical multipole order in general relativity.

\section{Application: Chaotic orbits for pulsating spheres in (vacuum) Schwarzschild spacetime}
\label{secchaos}

Our second application explores a different consequence of the same mechanism: once the hexadecapole force can be nonzero even in vacuum, time-dependent (pulsating) hexadecapole moments act as a periodic forcing on the orbital dynamics, potentially destroying integrability and generating chaotic layers. We focus on equatorial motion in Schwarzschild spacetime and ask whether periodic pulsations of a spherical, spinless body can trigger chaos near the separatrix associated with an unstable circular orbit.

The equation of motion is again Eq.~\eqref{eqmotionsimp}. We restrict to the equatorial plane $\theta=\pi/2$ and impose $u^\theta=0$, but we now allow nonzero angular momentum $L\neq 0$, so that $u^\mu$ and $v^\mu$ are no longer parallel in general. The nonzero components of the hexadecapole force $F_\mu$ are given in Appendix~\ref{ap:expressions}. Following the same procedure as in \cite{mosnaRodriguesVieira2022PRD}, one can reduce Eq.~\eqref{eqmotionsimp} to a two-dimensional nonautonomous system for the canonical variables $(r,p_r)$, valid to first order in the hexadecapole parameters:
\begin{align}
  \dot{r}&=f_1(r,p_r)+J_m(\tau)\,g_{1m}(r,p_r)+J_s(\tau)\,g_{1s}(r,p_r),
  \label{dotr}\\
  \dot{p}_r&=f_2(r,p_r)+J_m(\tau)\,g_{2m}(r,p_r)+J_s(\tau)\,g_{2s}(r,p_r).
  \label{dotpr}
\end{align}
The explicit expressions for $f_1,f_2$ and $g_{im},g_{is}$ are displayed in Appendix~\ref{ap:expressions}. With the conserved quantities $p_t$ and $p_\phi$ held fixed, a direct differentiation of those expressions gives
\begin{align}
  \partial_r f_1+\partial_{p_r}f_2&=0, \nonumber\\
  \partial_r g_{1m}+\partial_{p_r}g_{2m}&=0, \qquad
  \partial_r g_{1s}+\partial_{p_r}g_{2s}=0.
  \label{hamcheck}
\end{align}
Thus Eqs.~\eqref{dotr}--\eqref{dotpr} define a (time-dependent) Hamiltonian system in the canonical variables $(r,p_r)$, to the order retained. The unperturbed case $J_m=J_s=0$ corresponds to a point particle in Schwarzschild spacetime and is integrable; in particular it possesses an unstable fixed point $(r=r_{un},p_r=0)$ associated with the unstable circular orbit and the corresponding homoclinic orbit (separatrix) in the $(r,p_r)$ plane.

\subsection*{A simple pulsation model}

We now specify a simple periodic time dependence for the multipoles. For simplicity we take $J_s(\tau)=0$ and set
\begin{align}
  J_m(\tau)=J_0\big[1+\varepsilon\sin(\Omega\tau)\big],
  \qquad 0<\varepsilon<1,
  \label{oscillation}
\end{align}
where $J_0$ sets the (small) overall hexadecapole scale and $\varepsilon$ controls the oscillation amplitude. The nonzero mean keeps $J_m(\tau)$ nonnegative, as suggested by the mass-type integral in Eq.~\eqref{integralJ}, but no detailed internal matter model is assumed. The constant part $J_0$ produces an \emph{autonomous} (time-independent) hexadecapole correction to the point-particle dynamics, i.e.\ a mild deformation of the effective Hamiltonian and of the location of the separatrix. The genuinely nonintegrable ingredient is the oscillatory part $\propto \sin(\Omega\tau)$, which provides a time-periodic forcing and can split the stable and unstable manifolds associated with the separatrix. Since Melnikov's method diagnoses this splitting at leading order in the time-periodic perturbation, it is precisely the oscillatory component that matters for the onset of chaos.

\subsection*{Melnikov analysis and separatrix splitting}

Melnikov's method provides an efficient criterion for transverse intersections between the stable and unstable manifolds of a perturbed saddle in the Poincar\'e stroboscopic map \cite{Holmes1990}. Consider a Hamiltonian system
\begin{align}
  \dot{r}=&f_1(r,p_r)+\epsilon\,\lambda_1(r,p_r,\tau),\\
  \dot{p}_r=&f_2(r,p_r)+\epsilon\,\lambda_2(r,p_r,\tau),
\end{align}
where $\lambda_i$ are $2\pi/\Omega$-periodic in $\tau$ and $\epsilon$ is small. When the unperturbed system has a homoclinic orbit $(r(\tau),p_r(\tau))$ associated with an unstable fixed point, the leading-order distance between the perturbed stable and unstable manifolds (in a transverse direction) is proportional to the Melnikov function
\begin{align}
  M(\tau_0)=\int_{-\infty}^{\infty}\big(f_1\lambda_2-f_2\lambda_1\big)\big(r(\tau),p_r(\tau),\tau+\tau_0\big)\,d\tau,
\end{align}
evaluated along the unperturbed homoclinic orbit. If $M(\tau_0)$ has isolated simple zeros as a function of $\tau_0$, the manifolds intersect transversely, producing a homoclinic tangle and chaotic layers near the separatrix. If $M(\tau_0)$ vanishes identically, the first-order Melnikov criterion is inconclusive.

In our case, the time-periodic part of Eq.~\eqref{oscillation} yields a perturbation of the form $\epsilon\lambda_i=\big(J_0\varepsilon\sin(\Omega\tau)\big)g_{im}$. Up to an overall factor proportional to $J_0\varepsilon$ (which does not affect the location of the zeros), the resulting Melnikov function takes the same factorized form obtained in \cite{mosnaRodriguesVieira2022PRD}:
\begin{align}
  M(\tau_0)=2\cos(\Omega\tau_0)\,K(\Omega),
\end{align}
where $K(\Omega)$ is independent of $\tau_0$ and is given by
\begin{align}
  K(\Omega)=\int_{r_{un}}^{r_m}k(r)\sin\!\big[\Omega\tau(r)\big]\,dr.
\end{align}
Here $r_m$ denotes the outer radial turning point of the unperturbed homoclinic orbit. For Schwarzschild timelike geodesics parametrized by the unstable circular radius $r_{un}$, with $4M<r_{un}<6M$,
\begin{align}
  r_m=\frac{2Mr_{un}}{r_{un}-4M}.
\end{align}
Here $\tau(r)$ denotes the proper-time parametrization of the unperturbed Schwarzschild homoclinic orbit. The radial kernel for the hexadecapolar perturbation is
\begin{align}
  k(r)=&\frac{16 M^2}{15 r^{11}(r_{un}-3M)^2}\left(4Mr^2r_{un}(3r^2+r_{un}^2)\right.\nonumber\\
  &\left.-2r^4r_{un}^2-M^2(18r^4+12r^2r_{un}^2-5r_{un}^4)\right).
\end{align}

\begin{figure}[tbp]
	\includegraphics[width=\columnwidth]{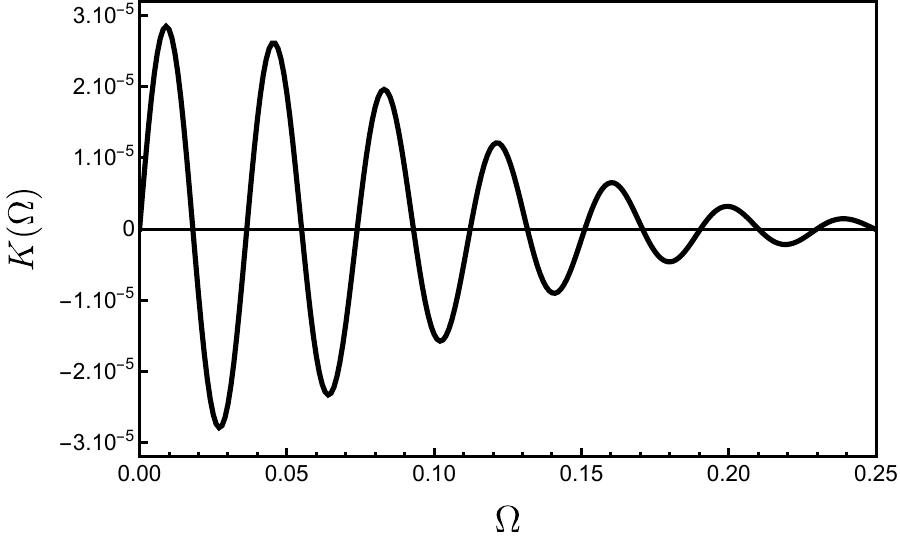} 
	\\ \hfill \\
	\includegraphics[width=\columnwidth]{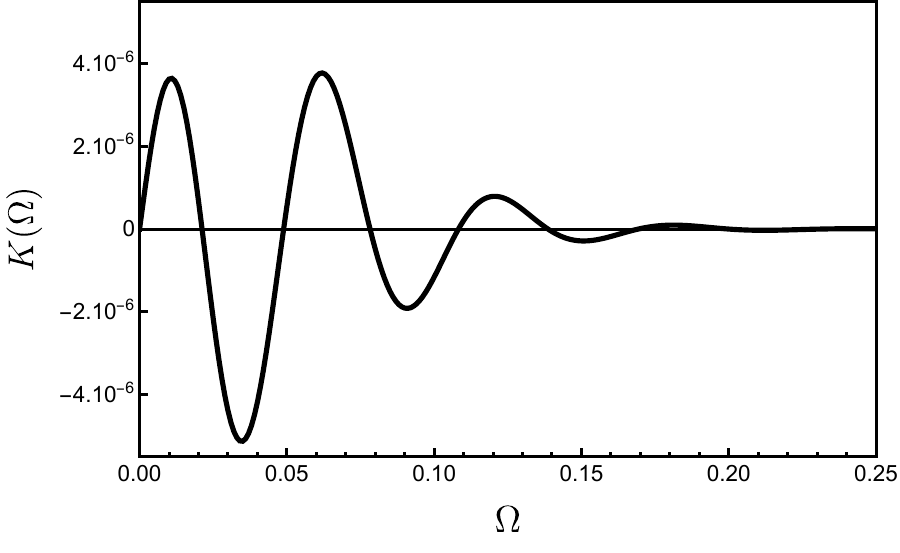}
	\caption{Plot of the function $K(\Omega)$ for $r_{un}=4.5 M$ (top) and $r_{un}=5 M$ (bottom). The nonidentically-zero oscillatory behavior shown here is typical of our scans; the zeros are isolated, so the Melnikov criterion predicts separatrix splitting for almost all driving frequencies, except at the isolated zeros of $K(\Omega)$.}
	\label{graphkomega}
\end{figure}

Since $\cos(\Omega\tau_0)$ has only isolated simple zeros, the existence of transverse homoclinic intersections is controlled by whether $K(\Omega)$ vanishes. In the representative cases shown in Fig.~\ref{graphkomega}, and in the broader numerical scans we performed over the relevant homoclinic-orbit parameter range, $K(\Omega)$ is not identically zero and its zeros occur only at isolated frequencies. Thus, in the usual generic sense, the Melnikov function has simple zeros for almost all driving frequencies $\Omega$, except at the isolated zeros of $K(\Omega)$, and the stable and unstable manifolds intersect transversely. This is a local statement in phase space: the splitting occurs near the homoclinic separatrix associated with the unstable circular orbit. In the black-hole problem the corresponding chaos is naturally transient, because nearby trajectories may eventually plunge or scatter rather than remain confined for all time.

This result complements the radial-infall calculation of Sec.~\ref{secfall}: both effects are direct consequences of the same breakdown of point-particle universality at the first nontrivial spherical multipole order in general relativity. While the constant hexadecapole moments already produce nongeodesic corrections in vacuum, time-dependent hexadecapole moments act as a periodic forcing that generically leads to separatrix splitting and chaos near the unstable circular orbit.

\section{Conclusion}
\label{conclusion}

In this work we revisited the universality of free fall for \emph{small spherical bodies} in vacuum by contrasting the Newtonian vacuum result with the relativistic dynamics of extended \emph{test} bodies in Dixon's multipole formalism. The Newtonian comparison (Sec.~\ref{secnewton} and Appendix~\ref{ap:newtexp}) highlights the key mechanism behind universality: in a source-free region the gravitational potential is harmonic, and spherical symmetry forces all higher multipole couplings to collapse into Laplacians, which vanish identically. A spherical body therefore follows exactly the same orbit as a point particle placed at its center of mass.

On the relativistic side, we formulated spherical symmetry within a fixed choice of center-of-mass worldline (Tulczyjew-Dixon SSC) and an $M$-transported tetrad, imposing $O(3)$ invariance in the instantaneous 3-space orthogonal to $p^\mu$ (Secs.~\ref{dixon} and~\ref{sec4pole}). Within this setting, at quadrupole order the torque vector vanishes identically for spherical bodies and a spinless configuration is dynamically consistent. Moreover, the quadrupole force for a spherical body reduces to Ricci-dependent terms [Eq.~\eqref{quadforcesph}], and therefore vanishes in any vacuum spacetime ($R_{\mu\nu}=0$). As a consequence, the center-of-mass worldline of a spinless spherical body is geodesic in \emph{any} Ricci-flat background, with no assumption of spacetime symmetry (including the Kerr metric). Spherical symmetry also implies that the octupole moment vanishes identically, so no deviation from geodesic motion appears at that order. We emphasize that this ``effacement-like'' statement is made strictly in the test-body regime with prescribed multipole moments, and should not be conflated with post-Newtonian effacement results for self-gravitating compact objects often discussed in connection with the strong equivalence principle.

The second main message is that this relativistic ``effacement-like'' behavior has a sharp limit. At the next nontrivial spherical order, the hexadecapole level, the torque vector remains identically zero, again allowing spinless motion, but the force generically does \emph{not} vanish even in vacuum. The structural reason is visible directly in the hexadecapole force [Eq.~\eqref{16force}]: besides terms involving derivatives of curvature, it contains contributions quadratic in the Riemann tensor. In Ricci-flat spacetimes these terms probe the Weyl curvature and are not forced to vanish by spherical symmetry alone; in particular, in vacuum ($R_{\mu\nu}=0$) the relevant curvature coupling is entirely Weyl-driven. Schwarzschild spacetime provides an explicit example where the spherical hexadecapole force is nonzero, identifying $n=4$ as the first multipole order at which point-particle universality can break for spherical bodies in vacuum general relativity.

We illustrated these general conclusions with two applications in Schwarzschild spacetime. First, for radial infall (Sec.~\ref{secfall}) we obtained a closed expression for the only nonzero component of the hexadecapole force, $F_r \propto M^2(8J_m-21J_s)/r^7$, and integrated the resulting dynamics to compute the leading correction to the proper time accumulated from a finite radius to the horizon for a trajectory compatible with release from rest at infinity. The correction has a definite sign set by the combination $h\propto(8J_m-21J_s)$; for simple matter models where the stress contribution is suppressed relative to the mass contribution, one expects $h>0$, implying a slightly longer infall proper time than for a point particle under the same conditions. Second, for equatorial motion with nonzero angular momentum (Sec.~\ref{secchaos}), we showed that time-dependent (pulsating) hexadecapole moments act as a periodic forcing on an otherwise integrable point-particle system. For the simple oscillation model used here, the periodic component generically splits the stable and unstable manifolds associated with the homoclinic orbit near the unstable circular orbit. Melnikov's method then predicts chaotic layers for almost all driving frequencies, except at the isolated values where the Melnikov amplitude $K(\Omega)$ vanishes.

Finally, it is important to stress the scope and limitations of the present analysis. All results are derived in the test-body regime and within the validity domain of Dixon's multipole expansion: the body is assumed small compared with the local curvature radius, backreaction is neglected, and the dynamics is truncated at a finite multipole order (with forces treated to leading order in the corresponding multipole parameters). Consequently, the effects discussed here are expected to be small in weak-field regions and become relevant only in sufficiently strong curvature or for sufficiently large (yet still perturbative) higher multipole moments. Within these limitations, the results provide a clear principle-level statement: in general relativity, spherical symmetry protects geodesic motion in vacuum only up to the quadrupole level, while at the first nontrivial spherical higher order ($n=4$) internal structure can affect the orbit even in a vacuum spacetime, with concrete consequences ranging from invariant timing corrections (infall) to nonintegrable and chaotic behavior under periodic internal forcing. This finite-order protection contrasts with Newtonian gravity, where the spherical vacuum cancellation holds at all multipole orders.

\paragraph*{Note added.}
While this manuscript was being finalized for submission, we became aware of
the newly posted preprint by Harte and Ramond
\cite{HarteRamond2026}. They show that only the trace-free part of a generic
octupole can affect motion in vacuum and point out that the analogous
decoupling of trace components need not persist at hexadecapole order. The
present work was completed independently and treats the complementary sector
of exactly spherical bodies: it carries out the $O(3)$-symmetric reduction
at hexadecapole order, establishes the dynamical consistency of the spinless
sector, and derives explicit nongeodesic dynamics for such bodies in
Schwarzschild spacetime.

\begin{acknowledgments}
R.A.M. was partially supported by Conselho Nacional de Desenvolvimento
Cient\'ifico e Tecnol\'ogico (CNPq, Brazil) under Grant No.~316780/2023-5 and by
Funda\c{c}\~ao de Amparo \`a Pesquisa do Estado de S\~ao Paulo (FAPESP) under
Grant No.~2024/00923-6.
The authors used \mbox{OpenAI} ChatGPT and \mbox{OpenAI} Codex tools as interactive
aids for exploratory discussion, editing, \mbox{LaTeX} preparation, and auxiliary
code/plot checks.  The authors remain fully responsible for this work.
\end{acknowledgments}

\appendix

\section{Multipole expansion of the Newtonian force and torque}
\label{ap:newtexp}

In this Appendix we provide a compact multipolar argument for the Newtonian result
discussed in Sec.~\ref{secnewton}: in a source-free region, a spherically symmetric body
experiences neither higher-multipole force corrections nor any torque.

Let $\Phi$ be a Newtonian gravitational potential generated by external sources.
In a source-free region it is harmonic,
\begin{equation}
\nabla^2 \Phi = \delta^{ij}\partial_i\partial_j \Phi = 0,
\end{equation}
and by commutativity of partial derivatives the same holds for any partial derivative of $\Phi$.

Consider an extended body with mass density $\rho(\mathbf{x},t)$, compact support
$\mathcal{R}(t)$, and center of mass $\mathbf{z}(t)$ defined by Eq.~\eqref{ncom}.
Define the separation vector components $r^i := x^i - z^i(t)$ and the (Cartesian) ``raw''
multipole moments
\begin{equation}
m^{i_1\cdots i_n}(t)
:= \int_{\mathcal{R}(t)} r^{i_1}\cdots r^{i_n}\,\rho(\mathbf{x},t)\,d^3\mathbf{x}.
\label{eq:rawmoments}
\end{equation}
Taylor-expanding $\partial_i\Phi(\mathbf{x})$ about $\mathbf{z}(t)$ and substituting into
the force definition Eq.~\eqref{nforce} yields
\begin{align}
F_i(t)
&= -\int_{\mathcal{R}(t)}\rho(\mathbf{x},t)\,\partial_i\Phi(\mathbf{x})\,d^3\mathbf{x}\nonumber\\
&= -m\,\partial_i\Phi(\mathbf{z}(t)) \nonumber\\
&-\sum_{n=2}^{\infty}\frac{1}{n!}\,m^{i_1\cdots i_n}(t)\,
\partial_i\partial_{i_1}\cdots\partial_{i_n}\Phi(\mathbf{z}(t)).
\label{eq:force_multipole}
\end{align}
The sum starts at $n=2$ because the dipole moment $m^{i}(t)$ vanishes by the definition of
$\mathbf{z}(t)$.

Similarly, substituting the same Taylor expansion into the torque definition Eq.~\eqref{ntorque}
gives
\begin{align}
N_i(t)
&= -\int_{\mathcal{R}(t)}\rho(\mathbf{x},t)\,\epsilon_{ijk}\,r^j\,\partial^k\Phi(\mathbf{x})\,d^3\mathbf{x}\nonumber\\
&= -\sum_{n=1}^{\infty}\frac{1}{n!}\,\epsilon_{ijk}\,m^{j j_1\cdots j_n}(t)\,
\partial^k\partial_{j_1}\cdots\partial_{j_n}\Phi(\mathbf{z}(t)),
\label{eq:torque_multipole}
\end{align}
where again the $n=0$ term drops out because $m^j(t)=0$.

\paragraph*{Isotropy of spherical moments.}
If the body is spherically symmetric, $\rho(\mathbf{x},t)=\rho(r,t)$ with $r=|\mathbf{r}|$, then there is
no preferred spatial direction in the rest frame of $\mathbf{z}(t)$, and the tensors
$m^{i_1\cdots i_n}$ are $O(3)$-invariant (isotropic). In particular, by parity one has
\begin{equation}
m^{i_1\cdots i_n}=0 \quad \text{for odd } n,
\end{equation}
and for even rank $n=2\ell$ isotropy implies that $m^{i_1\cdots i_{2\ell}}$ is proportional to
the unique $O(3)$-invariant totally symmetric tensor of that rank, namely a totally symmetrized
product of Kronecker deltas:
\begin{equation}
m^{i_1\cdots i_{2\ell}}
= A_{2\ell}\,\delta^{(i_1 i_2}\delta^{i_3 i_4}\cdots \delta^{i_{2\ell-1} i_{2\ell})}.
\label{eq:isotropic_moments}
\end{equation}
Equivalently, Eq.~\eqref{eq:isotropic_moments} represents the fully symmetric sum over all pairwise contractions of the indices. A convenient choice for the scalar
coefficient is
\begin{equation}
A_{2\ell}
= \frac{1}{(2\ell+1)!!}\int_{\mathcal{R}(t)} \rho(\mathbf{x},t)\,|\mathbf{r}|^{2\ell}\,d^3\mathbf{x},
\end{equation}
which reproduces, for example, $m^{ij}=\frac{1}{3}\delta^{ij}\int\rho r^2 d^3x$ and
$m^{ijkl}=\frac{1}{15}\delta^{(ij}\delta^{kl)}\int\rho r^4 d^3x$.

\paragraph*{Force: all higher multipoles vanish in vacuum.}
Inserting Eq.~\eqref{eq:isotropic_moments} into the $n\ge 2$ terms of
Eq.~\eqref{eq:force_multipole}, each Kronecker delta contracts a pair of partial derivatives.
By commutativity of partial derivatives, this produces Laplacians acting on derivatives of $\Phi$,
e.g.\ terms of the form
\begin{align}
\delta^{ab}\partial_a\partial_b\left(\partial_{c_1}\cdots\partial_{c_k}\Phi\right)
=& \nabla^2\left(\partial_{c_1}\cdots\partial_{c_k}\Phi\right) \nonumber\\
=& \partial_{c_1}\cdots\partial_{c_k}\left(\nabla^2\Phi\right)=0,
\end{align}
in a source-free region. Therefore every multipolar correction in the sum in
Eq.~\eqref{eq:force_multipole} vanishes, and the total force reduces to the monopole term,
\begin{equation}
F_i(t) = -m\,\partial_i\Phi(\mathbf{z}(t)).
\end{equation}

\paragraph*{Torque: vanishing at each multipole order.}
For the torque, Eq.~\eqref{eq:torque_multipole}, isotropy implies that $m^{j j_1\cdots j_n}$ is
a sum of products of Kronecker deltas pairing its indices. In every such term, the index $j$
must be paired with one of the $j_a$ indices, producing a factor $\delta^{j j_a}$ and hence a
second derivative $\partial^k\partial^{j}$ acting on $\Phi$. Since $\partial^k\partial^{j}$ is
symmetric in $(j,k)$, its contraction with $\epsilon_{ijk}$ vanishes. Consequently, each term in
the sum in Eq.~\eqref{eq:torque_multipole} is zero, and
\begin{equation}
N_i(t)=0
\end{equation}
order by order in the multipole expansion. Note that this last conclusion does not require $\Phi$ to be
harmonic; it follows from spherical symmetry alone, consistently with the main-text argument in
Sec.~\ref{secnewton}.

\section{Centroid convention and spherical symmetry}
\label{app:ssc_spherical}

In a multipolar description, the statement that a body is ``spherical'' is not
a statement about a list of components alone. The multipole moments are defined
relative to a representative worldline, just as Newtonian multipole moments are
defined relative to an origin. Changing the representative worldline generally
mixes the multipole hierarchy. In relativity one must also specify the local
spatial sections in which the moments are decomposed.

For definiteness, in the main text we use the Tulczyjew-Dixon
center-of-mass convention, $S^{\mu\nu}p_\nu=0$. Thus ``spherical''
means $O(3)$ invariance in the 3-space orthogonal to $p^\mu$, with the
components represented in an $M$-transported tetrad. This convention fixes
the bookkeeping used throughout the paper. Other centroid prescriptions can
of course be used, but then the corresponding notion of isotropy must be
formulated in their own spatial sections; the resulting multipole
components need not agree term by term with ours in a finite truncation.

This convention-dependence is not the origin of the quadrupolar Ricci-flat
cancellation. At quadrupole order the reason is algebraic. If an isotropic
quadrupole is built from any unit timelike direction $n^\mu$, with spatial
projector $h_n^{\mu\nu}=g^{\mu\nu}+n^\mu n^\nu$, it has the same trace
structure as Eq.~\eqref{jquad_h}. These traces project the Riemann tensor only
onto Ricci contractions; the trace-free Weyl part drops out. Hence the
quadrupolar cancellation in Ricci-flat backgrounds is not a peculiarity of the
Tulczyjew-Dixon prescription.

At higher multipolar order this simple trace argument no longer applies.
In particular, the hexadecapole force contains curvature-squared terms, and
isotropic higher moments can contract with Weyl curvature even when
$R_{\mu\nu}=0$. Our explicit hexadecapole formulae are therefore stated in
the Tulczyjew-Dixon bookkeeping. A change of centroid convention may reshuffle
terms among multipolar orders in a finite truncation, but it does not provide
an algebraic reason for restoring vacuum geodesic universality beyond
quadrupole order.

\par\vspace{0.5\baselineskip}
\section{Explicit Schwarzschild expressions for the hexadecapole force}
\label{ap:expressions}

For the equatorial Schwarzschild orbits considered in Sec.~\ref{secchaos},
the spinless hexadecapole equation of motion \eqref{eqmotionsimp} involves
only two nonzero components of the force, $F_r$ and $F_\phi$. In terms of
the conserved quantities $p_t$ and $p_\phi$, obtained from
Eq.~\eqref{conserved}, these components are as follows. We also list the
coefficient functions $f_1,f_2,g_{1m},g_{1s},g_{2m},g_{2s}$ entering the
reduced system \eqref{dotr}--\eqref{dotpr}.

\onecolumngrid

\begin{align}
	F_r =& \frac{4 M^2}{15 A^4 r^7} \Big(\left(8 J_m-21 J_s\right) \left(\left(r (r-2 M)^2 p_r^2-r^3 p_t^2\right){}^2+(r-2 M)^2 p_{\phi }^4\right)\nonumber\\
	&+2 \left(14 J_m-3 J_s\right) r (r-2 M) p_{\phi }^2 \left((r-2 M)^2 p_r^2-r^2 p_t^2\right)\Big),\\
	F_\phi =& -\frac{16 M^2}{15 A^4 r^6} \left(J_m+3 J_s\right) (r-2 M)^2 p_r p_{\phi } \left(r^3 p_t^2 - r (r-2 M)^2 p_r^2 +(r-2 M) p_{\phi }^2\right),\\
	f_1 =& \frac{1}{A}(r-2 M)^{3/2} p_r,\\
	g_{1m} =& -\frac{16 M^2}{15 A^6 r^5} (r-2 M)^3 p_r p_{\phi }^2 \left(r^3 p_t^2 -r (r-2 M)^2 p_r^2 + (r-2 M) p_{\phi }^2 \right),\\
	g_{1s} =& 3 g_{1m},\\
	f_2 =& \frac{(r-2 M)^2 p_{\phi }^2-M r \left((r-2 M)^2 p_r^2+r^2 p_t^2\right)}{A r^2 (r-2 M)^{3/2}},\\
	g_{2m} =& -\frac{16 M^2}{15 A^6 r^7}\Big( r^6 (10 r-21 M) p_t^4 p_{\phi }^2-r^3 (r-2 M) (8 r-15 M) p_t^2 p_{\phi }^4+2 r^3 (r-2 M)^6 p_r^6\nonumber\\
	&+r (r-2 M)^2 p_r^2 \left(6 r^6 p_t^4-20 r^3 (r-2 M) p_t^2 p_{\phi }^2+(r-2 M) (8 r-17 M) p_{\phi }^4\right)\nonumber\\
	&-r^2 (r-2 M)^4 p_r^4 \left(6 r^3 p_t^2-(10 r-19 M) p_{\phi }^2\right)+2 (r-2 M)^3 p_{\phi }^6-2 r^9 p_t^6 \Big),\\
	g_{2s} =& \frac{4 M^2}{5 A^6 r^7} \Big(r^6 (5 r-6 M) p_t^4 p_{\phi }^2-r^3 (r-2 M) (13 r-30 M) p_t^2 p_{\phi }^4+7 r^3 (r-2 M)^6 p_r^6\nonumber\\
	&+r (r-2 M)^2 p_r^2 \left(21 r^6 p_t^4-10 r^3 (r-2 M) p_t^2 p_{\phi }^2+(r-2 M) (13 r-22 M) p_{\phi }^4\right)\nonumber\\
	&-r^2 (r-2 M)^4 p_r^4 \left(21 r^3 p_t^2-(5 r-14 M) p_{\phi }^2\right)+7 (r-2 M)^3 p_{\phi }^6-7 r^9 p_t^6\Big),
\end{align}
where
\begin{align}
	A = \sqrt{r^3 p_t^2-(r-2 M) \left[r (r-2 M) p_r^2+p_{\phi }^2\right]}.
	\label{eq:A_schw}
\end{align}

The square root in Eq.~\eqref{eq:A_schw} is taken on the positive branch outside
the horizon. As a useful check, in the radial limit $p_\phi=0$ one has
$F_\phi=0$ and
\begin{equation}
\left[r(r-2M)^2p_r^2-r^3p_t^2\right]^2=A^4,
\end{equation}
so that the first equation above reduces to
\begin{equation}
F_r=\frac{4M^2}{15r^7}(8J_m-21J_s),
\end{equation}
which is the radial force used in Sec.~\ref{secfall}. For $p_\phi\neq0$, the
same expressions give the coefficients of the nonautonomous system
\eqref{dotr}--\eqref{dotpr} used in Sec.~\ref{secchaos} and therefore provide
the explicit input for the Melnikov analysis.

\twocolumngrid

\end{document}